\documentclass[letter]{aa}
\usepackage{graphicx}
\usepackage{txfonts}
\usepackage{subfig}

\def\kms{km\,s$^{-1}$}
\def\zgt6{\hbox{{\it z}$\,>\,$6}}

\def\lsim{\mathrel{\rlap{\lower 3pt \hbox{$\sim$}} \raise 2.0pt \hbox{$<$}}}
\def\gsim{\mathrel{\rlap{\lower 3pt \hbox{$\sim$}} \raise 2.0pt \hbox{$>$}}}

\begin{document}

\authorrunning{Castignani et al.}
\titlerunning{Environmental processing at $z=1.7$}

\title{Environmental processing in cluster core galaxies at z=1.7}

\author{G. Castignani
          \inst{1}\fnmsep\thanks{e-mail: gianluca.castignani@epfl.ch}
          \and
          F. Combes\inst{2,3}
          \and
          P. Salom\'e\inst{2}
          }

   \institute{Laboratoire d'astrophysique, \'{E}cole Polytechnique F\'{e}d\'{e}rale de Lausanne (EPFL), Observatoire de Sauverny, 1290 Versoix, Switzerland
              \and
         Sorbonne Universit\'{e}, Observatoire de Paris, Universit\'{e} PSL, CNRS, LERMA, F-75014, Paris, France
               \and
             Coll\`{e}ge de France, 11 Place Marcelin Berthelot, 75231 Paris, France
             }
\date{Received November 20, 2019; Accepted February 14, 2020}

\abstract{Today, the brightest cluster galaxies (BCGs) are passive and very massive galaxies at the center of their clusters, and they
still accrete mass through swallowing companions and gas from cooling flows. However their formation history is not well known. We report CO(4$\rightarrow$3) and continuum map observations of the SpARCS1049+56 BCG at $z=1.709$, one of the most distant known BCGs. Our observations yield $M_{{\rm H}_2}<1.1\times10^{10}M_\odot$ for the BCG; while in CO(4$\rightarrow$3), we detect two gas-rich companions { at the northeast and southeast} of the BCG, within 20~kpc, with ${L^\prime_{\rm CO(4\rightarrow3)}=(5.8\pm0.6)\times10^{9}}$~K~km~s$^{-1}$~pc$^2$ and ${(7.4\pm0.7)\times10^{9}}$~K~km~s$^{-1}$~pc$^2$, respectively. 
The northern companion is associated with a pair of merging cluster galaxies, while the { southern one} shows a southern tail in CO(4$\rightarrow$3), which was also detected in continuum, and we suggest it to be the most distant jellyfish galaxy for which ram pressure stripping is effectively able to strip off its dense molecular gas. This study probes the presence of rare gas-rich systems
in the very central region of a distant cluster core, which will potentially merge into the BCG itself. Currently, we may thus be seeing the reversal of the star formation versus density relation at play in the distant universe.
This is the first time the assembly of high-$z$ progenitors of our local BCGs can be studied in such great detail.}


\maketitle

\section{Introduction}
The brightest cluster galaxies (BCGs) are  excellent laboratories to study the effect of a dense galaxy cluster environment on galaxy evolution. 
BCGs have exceptional masses and luminosities. In the local Universe, they are indeed commonly associated with passively evolving massive ellipticals of cD type, which often host radio galaxies \citep{Zirbel1996}, and are located at the center of the cluster cores \citep{Lauer2014}.

Present day star-forming BCGs ($>40~M_\odot/$yr), such as the famous Perseus~A and Cygnus~A \citep{FraserMcKelvie2014}, have been commonly found in cool-core clusters \citep{Reynolds_Fabian1996,Fabian2006}. This suggests that large-scale flows of gas regulate their star formation and are { often moderated or suppressed by active galactic nuclei (AGN) feedback}.
Some studies of local and intermediate-$z$ BCGs have indeed found molecular gas reservoirs to fuel the star formation \citep{Edge2001,Salome_Combes2003,Hamer2012,McDonald2014, McNamara2014,Russell2014,Tremblay2016,Fogarty2019} as well as flows of cold gas, which is often in filaments \citep{Salome2006,Olivares2019,Russell2019}.

Little is known about the formation history of BCGs. They are commonly believed to evolve via the dry accretion of satellite galaxies \citep{Collins2009,Stott2011}, which is consistent with recent studies proposing that BCGs grow by at factor of $\sim2$ in stellar mass since $z\sim1$ \citep{Lidman2012,Zhang2016}.  However, most of the BCG stellar mass is assembled at higher redshifts ($z\sim3-5$) in smaller sources, which  are then later swallowed by the BCG \citep{DeLucia_Blaizot2007}. 

Recent studies have even proposed that the star formation in distant $z\gtrsim1$ BCGs is driven by gas-rich major mergers, resulting in high star formation and large reservoirs of molecular gas in BCGs at high $z\geq1$
\citep{McDonald2016,Webb2015a,Webb2015b,Bonaventura2017}, which is a scenario where star formation is fed by rapid gas deposition in distant BCGs and slow cooling flows at low-$z$.  In this work, we tested this scenario at $z = 1.7$ by mapping the sturbursting core and the BCG \citep[${\rm SFR} = 860\pm130~M_\odot$/yr,][]{Webb2015b} of the distant cluster SpARCS1049+56 in CO.

SpARCS1049+56 is one of the most distant and massive clusters 
\citep[e.g.,][]{Stanford2012,Brodwin2012,Newman2014,Gobat2013} with { weak-lensing and richness-based (estimated within 500~kpc) masses of $(3.5\pm1.2)\times10^{14}~M_\odot$ and $(3.8\pm1.2)\times10^{14}~M_\odot$, respectively \citep{Andreon_Congdon2014,Webb2015b,Finner2020}, which are consistent with each other.}
\citet{Webb2017} recently detected CO(2$\rightarrow$1) with the Large Millimeter Telescope (LMT), which is associated with a large reservoir ($\sim10^{11}~M_\odot$) of molecular gas that is possibly associated with the BCG and/or its nearby cluster core companions within the 25~arcsec
diameter (i.e., $\sim200$ kpc at $z = 1.7$) beam of the LMT.
{By spatially resolving, both in continuum and in CO, the BCG from each of its nearby companions separately, the present study} the present study is one of the first to probe both molecular gas and dust reservoirs of $z\gtrsim1$ BCGs in detail \citep{Emonts2013,Emonts2016,Coogan2018,Strazzullo2018,Castignani2019}. We thus aim to unveil the mechanisms at play for the assembly and the star formation fueling of the most distant BCGs.

Throughout this work, we adopt a flat $\Lambda \rm CDM$ cosmology with matter density $\Omega_{\rm m} = 0.30$, dark energy density $\Omega_{\Lambda} = 0.70,$ and Hubble constant $h=H_0/100\, \rm km\,s^{-1}\,Mpc^{-1} = 0.70$ \citep[see however,][]{PlanckCollaborationVI2018,Riess2019}. The luminosity distance at the redshift of the BCG is 12,811~Mpc { (i.e., $1'' = 8.46$~kpc).} The paper is structured as follows. In Sect.~\ref{sec:obs} we describe our NOEMA observations and data reduction; in Sect.~\ref{sec:results} we show our results. In Sect.~\ref{sec:discussion_conclusions} we discuss our results and draw our conclusions. In Appendix~\ref{app:formulas}, we report the formulas adopted to estimate H$_2$ and the dust masses.


\begin{figure*}
\begin{center}
 \subfloat[]{\includegraphics[width=0.26\textwidth]{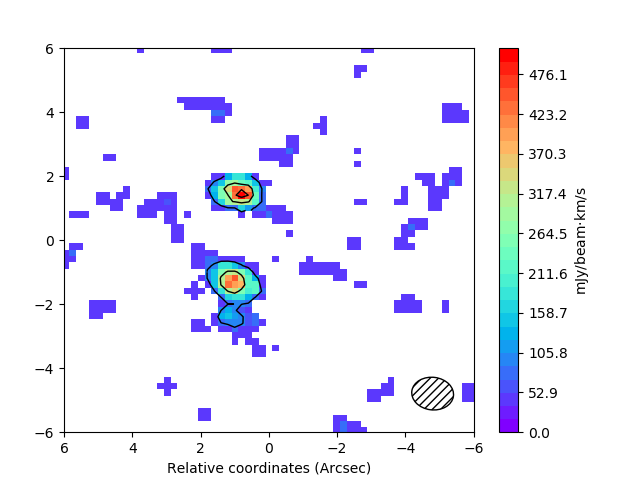}}
 \subfloat[]{\includegraphics[width=0.26\textwidth]{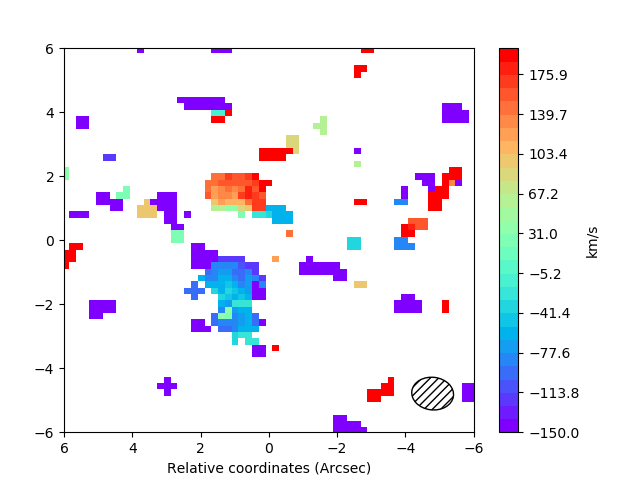}}
 \subfloat[]{\includegraphics[width=0.26\textwidth]{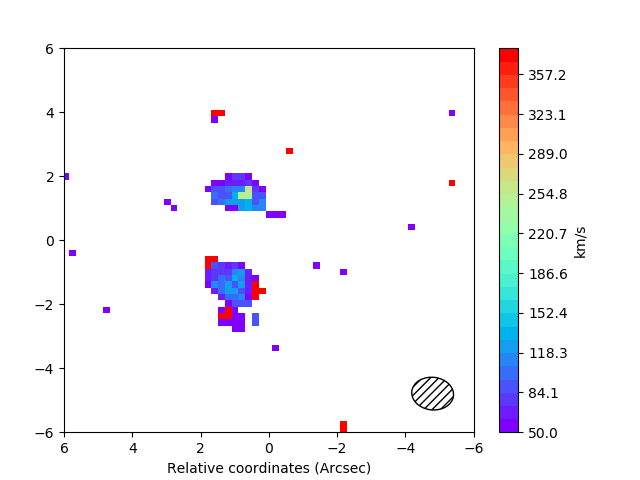}}
 \subfloat[]{\includegraphics[width=0.26\textwidth]{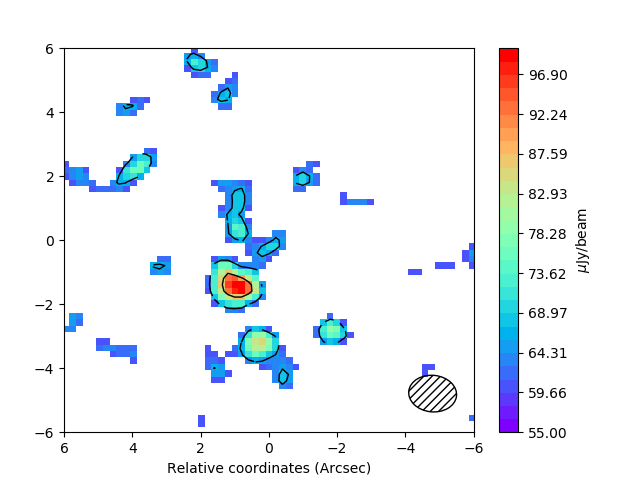}}
 \end{center}
 \caption{NOEMA CO(4$\rightarrow$3) observations. (a) Intensity, (b) velocity, and (c) velocity dispersion maps derived from the moments 0, 1, and 2, respectively. Panel~(d) shows the continuum map. Coordinates are reported as angular separations from the BCG. For each map, the velocity range that is considered corresponds to the velocity support of the CO(4$\rightarrow$3) line.  Solid contour levels are superimposed to both the intensity and continuum maps. They correspond to significance levels of 3, 10, and 15$\sigma$ (panel~a) as well as 3 and 4$\sigma$ (panel~d).  
 The dashed ellipses at the bottom right corner of each panel show the beam size.}
\label{fig:maps}
\end{figure*}

\begin{figure}
\begin{center}
 \includegraphics[width=0.5\textwidth]{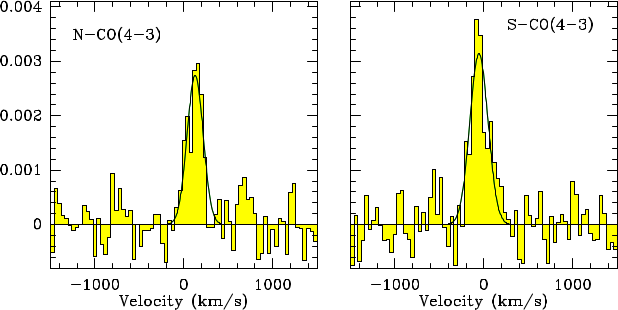}

 \includegraphics[width=0.25\textwidth]{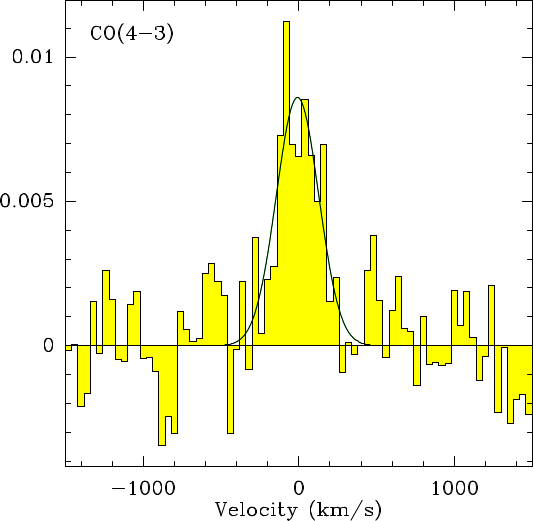}
\end{center}
 \caption{{Baseline subtracted spectra of northern (top left) and southern (top right) sources, and the system of the two (bottom).} 
 In the panels, the solid line shows the Gaussian best fit to the CO(4$\rightarrow$3) line. The vertical scale is in Jy.}
\label{fig:CO43spectrum}
\end{figure}

\begin{figure*}
\begin{center}
 \includegraphics[width=1.0\textwidth]{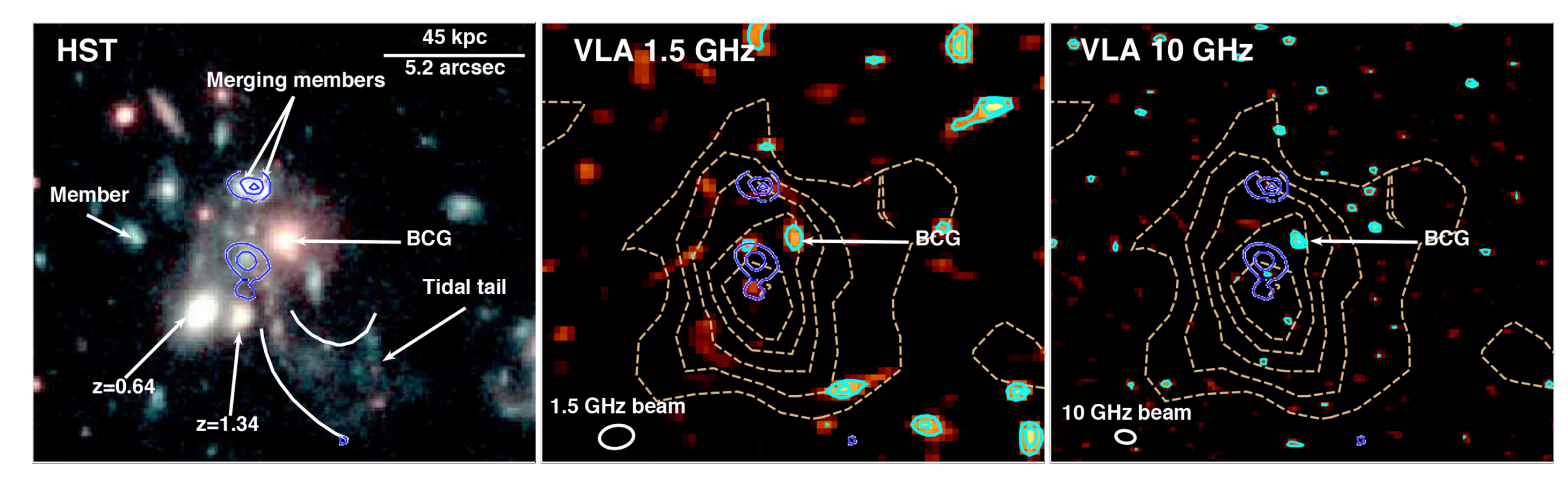}
\end{center}
 \caption{{\it HST} (left; RGB image: the F105W filter is in blue and green, F160W in red), VLA 1.5~GHz (center), and VLA 10~GHz (right) images of the BCG and its surroundings taken from \citet{Trudeau2019}. North is up, east is left. The CO(4$\rightarrow$3) contours are overplotted in blue, as in Fig.\ref{fig:maps}a. Small shifts of 0.3 and 0.6~arcsec toward the east and north, respectively, are applied to the NOEMA contours, which is consistent with GCS catalog uncertainties in the absolute astrometry used by {\it HST}. In the central and right panels, the radio contours (3, 4, and 5$\sigma$ levels) are in cyan and the {\it Spitzer} MIPS 24~$\mu$m contours are shown in beige.}
 \label{fig:Trudeau2019}
\end{figure*}

\section{NOEMA 1.8~mm observations and data reduction}\label{sec:obs}

We observed the SpARCS1049+56 BCG with NOEMA for four days in November 2018
(4, 15, 16, and 24 November) in C configuration using ten antennas, as part of the program S18CL (PI:~Combes). The total time spent on-source is 14.2h. We set the phase center of the observations equal to the BCG coordinates R.A.~=~$10^h49^m22.6^s$ and Dec.~=~$56^\circ40^\prime32.5^{\prime\prime}$.
We also adopted a tuning frequency of 170.2~GHz, which corresponds to the observer frame frequency of the redshifted CO(4$\rightarrow$3) line, given that our target BCG is at $z=1.709$ \citep{Webb2015b,Webb2017}.

The baseline range was 20--320\,m. The program was executed in good to typical winter weather conditions, with a system temperature of $T_{\rm sys}=150-250$\,K  and a precipitable water vapor column of $\sim2-3$\,mm and $\sim5-7$\,mm . We used the new PolyFix correlator, which covers a total bandwidth of 15.5~GHz split between the lower and upper side band, covering the ranges between 154.2-158.3~GHz and 169.7-173.8~GHz, respectively. The sources 1125+596, 0954+556 were used as phase and amplitude calibrators, while 3C84 and 3C279 served as bandwidth and flux calibrators. We reduced the data using the \textsf{clic} package from the \textsf{GILDAS} software. 
The final ($u$,$v$) tables correspond to 20.3\,hr of integration (14.2h on source).

We imaged the visibilities using the software \textsf{mapping} of \textsf{GILDAS}. At the tuning frequency of 170.2~GHz, the half power primary beam width is $29.6''$. We adopted natural weighting, yielding a synthesized beam of $1.2''\times1.0''$ with PA=82$^\circ$. We then rebinned the spectral axis at a resolution of 20\,\kms{} ($\sim11.4$\,MHz at 170.2 GHz). The resulting root mean square (rms) is 0.55\,mJy\,beam$^{-1}$.

\section{Results}\label{sec:results}
\subsection{CO(4$\rightarrow$3) emission line sources}
Our upper side band NOEMA observations are sensitive to the CO(4$\rightarrow$3) line from cluster members within a 16~arcsec radius (i.e., 135~kpc) from the BCG and in the redshift range between $z=1.65-1.72$, which corresponds to velocities $\sim\pm2,000$~km~s$^{-1}$ with respect to the BCG. 
Surprisingly, no CO(4$\rightarrow$3) emission line was found to be associated with the BCG.
However, by inspecting the NOEMA data cube, we detected two sources in the upper side band in the field of the BCG at { (R.A.;Dec.) = ($10^h49^m22.60^s;56^\circ40^\prime33.9^{\prime\prime}$) and ($10^h49^m22.63^s;56^\circ40^\prime31.1^{\prime\prime}$)}. No emission line detection was found in the lower side band, which is sensitive to CO(4$\rightarrow$3) from sources within $z = 1.91-1.99$. In Fig.~\ref{fig:maps} we show  moment 0, 1, and 2 maps. The intensity map shows the two detected sources, which are about $\pm$2~arcsec to the northeast and southeast from the BCG. They also show a clear separation in velocity, being located at $\sim{+130}$~km~s$^{-1}$ and $\sim-60$~km~s$^{-1}$ with respect to the BCG, respectively.
The moment 2 map shows similar extensions of $\sim$150-250~km~s$^{-1}$ in CO(4$\rightarrow$3) for the two sources along the spectral axis. Tentative evidence for line broadening was found from the periphery down to the center of the northern source, while an opposite trend was observed for the southern companion.
The two sources remain unresolved by our NOEMA observations. However, the southern source shows evidence for a southern extension, which is visible from the intensity map. The velocity map also indicates that this southern tail may be receding with a velocity $\gtrsim100$~km~s$^{-1}$ with respect to the main source, while the recession seems to disappear at the very end of the tail. The velocity gradient suggests that the southern extension is associated with a bent tail of molecular gas. { 
Similarly, the southern part of the northern CO emitting source seems to approach us  with a relative velocity $\lesssim100$~km~s$^{-1}$ with respect to the main component. The velocity gradients are overall small, suggesting that both northern and southern CO companions of the BCG are seen almost face-on.}

\begin{table*}[htb]
\begin{center}
\begin{tabular}{lccccccc}
\hline
Component & $S_{\rm CO(4\rightarrow3)}$ & FWHM & $L^\prime_{\rm CO(4\rightarrow3)}$ &  $z_{\rm CO(4\rightarrow3)}$ \\
 & (Jy~km~s$^{-1}$) & (km~s$^{-1}$) & ($10^{9}$~K~km~s$^{-1}$~pc$^2$) &   \\

\hline
North & 0.62$\pm$0.06 & 210$\pm$25 & 5.8$\pm$0.6 & 1.7102$\pm$0.0001 \\ 
South  & 0.80$\pm$0.08 & 235$\pm$31 & 7.4$\pm$0.7 & 1.7085$\pm$0.0001 \\ 
Total & 2.94$\pm$0.31 & 319$\pm$39 & 27.2$\pm$2.9  & 1.7089$\pm$0.0002  \\ 
\hline
 \end{tabular}
\end{center}
 \caption{CO(4$\rightarrow$3) results for the northern and southern components and the system of the two, denoted as {\it Total}.} 
\label{tab:COresults}
\end{table*}

%
%
%

\subsection{Molecular gas mass and excitation ratio of the BCG companions}\label{sec:result_MH2}
Figure~\ref{fig:CO43spectrum} displays the CO(4$\rightarrow$3) spectrum for both the northern and southern companions of the BCG, {which are considered separately, and for the system of the two, which are considered altogether, within the $\sim10$~arcsec$^2$ region that encompasses the two sources.} Our Gaussian fit yields an estimated velocity integrated flux ${S_{\rm CO(4\rightarrow3)}\Delta v=(2.9\pm0.3)}$~Jy~km~s$^{-1}$, a full width at half maximum ${{\rm FWHM}=(319\pm39)}$~km~s$^{-1}$, and a redshift of ${1.7089\pm0.0002}$ for the system of the two CO emitting sources. The redshift is fully consistent with $z= 1.7091\pm0.0004$, which is inferred by \citet{Webb2017} with CO(2$\rightarrow$1) Large Millimiter Telescope (LMT) observations of the BCG system, which was unresolved by the LMT with a 25~arcsec beam.  


By using Eq.~\ref{eq:LpCO}, we also estimated a total velocity integrated luminosity of ${L^\prime_{\rm CO(4\rightarrow3)}=(2.7\pm0.3)\times10^{10}}$~K~km~s$^{-1}$~pc$^2$, which we compared to $L^\prime_{\rm CO(2\rightarrow1)}=(1.16\pm0.10)\times10^{11}$~K~km~s$^{-1}$~pc$^2$ found by \citet{Webb2017} with the LMT. The ratio between the two yields an excitation ratio of $r_{42} =L^\prime_{\rm CO(4\rightarrow3)}/L^\prime_{\rm CO(2\rightarrow1)}={0.23\pm0.03}$, which is lower than the value $\sim0.5,  $ which is typically found for distant star-forming galaxies \citep{Carilli_Walter2013}. \citet{Webb2017} also find that CO(2$\rightarrow$1) ${\rm FWHM} = (569\pm63)$~km~s$^{-1}$, which is a factor of $\sim1.8$ higher than what we find with our NOEMA observation, which is typical of massive star-forming galaxies.

{It is likely that the CO(2$\rightarrow$1) emission reported by the authors is due to a number of cluster members within the LMT beam of $25^{\prime\prime}$, which we did not detect, individually, with the smaller beam of $\sim1^{\prime\prime}$ of our observations. }
In this case, the $r_{42}$ excitation ratio for the BCG companions would be in better agreement with the typical values from the literature.

{ We estimated the total H$_2$ gas mass by using Eq.~\ref{eq:MH2} as well as by assuming a Galactic CO-to-H$_2$ conversion factor $\alpha_{\rm CO}=4.36~M_\odot~({\rm K}~{\rm km}~{\rm s}^{-1}~{\rm pc}^2)^{-1}$, an excitation ratio $r_{21}=0.85$ \citep{Carilli_Walter2013}, and $L^\prime_{\rm CO(2\rightarrow1)}$ by \citet{Webb2017}, which was rescaled by a factor of $1/1.8$. This yields $M_{{\rm H}_2}=(3.3\pm0.3)\times10^{11}~M_\odot$, a factor of $\sim3$ higher than what \citet{Webb2017} report. { We stress that the authors adopted a lower $\alpha_{\rm CO}=0.8~M_\odot~({\rm K}~{\rm km}~{\rm s}^{-1}~{\rm pc}^2)^{-1}$ than the Galactic one used in this work. Therefore, the factor of $(4.36/1.8)/0.8\sim3$ difference between the two mass estimates is ultimately due to the different CO-to-H$_2$ conversions adopted in the two studies and to the rescaling factor used in this work.}

The relative contribution of the northern and southern BCG companions to the total CO(4$\rightarrow$3) flux is found to be {$\sim20\%$ and $\sim30\%$, which imply H$_2$ masses of $\sim7\times10^{10}~M_\odot$ and $\sim1\times10^{11}~M_\odot$, respectively. The sum of the CO(4$\rightarrow$3) emission from the northern and southern components accounts for $\sim50\%$ of the total. This suggests that the remaining $\sim50\%$ is likely due to diffuse emission from gas that has been stripped from cluster core galaxies. In Table~\ref{tab:COresults} we summarize our results.}}


\subsection{Continuum emission at 1.8~mm and dust masses of the BCG companions}
We further investigate the presence of submillimiter sources in
the field of the BCG. By using the entire lower side band, we looked for possible continuum emission with an rms$=22\mu$Jy. Figure~\ref{fig:maps}d displays the corresponding continuum map in which two detections are at $\sim(3-4)\sigma$, which are associated with the southern companion of the BCG and its southern tail, with an associated total flux of $\sim0.2$~mJy. The continuum map does not show any detection that is either associated with the northern companion of the BCG or with the BCG itself. Nevertheless, NOEMA observations allowed us to set a robust 3$\sigma$ upper limit of 66$\mu$Jy to the dust continuum of the BCG at 1.8~mm in the observer frame. Remarkably, this upper limit is in agreement with the expected value $\sim0.1$~mJy, which was inferred using the spectral energy distribution modeling by \citet{Trudeau2019} that is based on low-radio frequency and far-infrared data-points.

We used the continuum map to estimate a dust mass $M_d=(2.2\pm1.1)\times10^8~M_\odot$ for the southern companion of the BCG and its tail, which is similar to those previously found for distant star-forming galaxies \citep{Berta2016}. 
We report a fiducial $50\%$ error in $M_d$ in order to take the uncertainties in the continuum flux and continuum to dust mass conversion into account. 
For the BCG, we set a $3\sigma$ upper limit of $M_d<7.2\times10^7~M_\odot$. For our estimates, we used Eq.~\ref{eq:Mdust} which is based on previous studies by \citet{Scoville2014,Scoville2016} on distant star-forming galaxies. 
We adopted a dust temperature equal to  $T_d=(44\pm5)$~K, which was inferred by \citet{Webb2015b} by modeling the infrared spectral energy distribution of the unresolved BCG system. The {\it Spitzer} MIPS 24~$\mu$m contours that are overplotted in the center and right panels of Fig.~\ref{fig:Trudeau2019} indeed show that the BCG and their nearby companions are unresolved in the far-infrared.

%
\subsection{Molecular gas mass reservoir of the BCG}
Neither CO(4$\rightarrow$3) nor continuum emission was found at the location of the BCG. However, by using Eqs.~(\ref{eq:LpCO}, \ref{eq:MH2}), we set a $3\sigma$ upper limit for the H$_2$ molecular gas mass of the BCG that is equal to $M_{\rm H_2}<1.1\times10^{10}~M_\odot$, corresponding to $L^\prime_{\rm CO(4\rightarrow3)}<1.2\times10^9$~K~km~s$^{-1}$~pc$^2$. 
We adopted a resolution of 300~km~s$^{-1}$, a 
Galactic CO-to-H$_2$ conversion factor $\alpha_{\rm CO}=4.36~M_\odot~({\rm K}~{\rm km}~{\rm s}^{-1}~{\rm pc}^2)^{-1}$, and an excitation ratio $r_{41}=0.46$ \citep{Carilli_Walter2013}.
By using Eq.~\ref{eq:Mmol_continuum} and the upper limit of the BCG continuum, we also set a 3$\sigma$ upper limit $M_{\rm H_2}<1.7\times10^{10}~M_\odot$, which is consistent with what was obtained from the nondetection in CO(4$\rightarrow$3).


\section{Discussion and conclusions}\label{sec:discussion_conclusions}

Figure~\ref{fig:Trudeau2019} displays the {\it Hubble Space Telescope} ({\it HST}) and Very Large Array (VLA) radio images of the BCG and its nearby companions. In the left panel, the cluster members are highlighted and they have been spectroscopically confirmed by \citet{Webb2015b}. Interestingly, the northern CO emitting source is coincident with a pair of merging cluster members. With this work, we also spectroscopically confirm the southern CO emitting source as a new cluster galaxy, with a clear {\it HST} counterpart. Interestingly, this source is also coincident with the {\it Spitzer} MIPS peak at 24~$\mu$m in the observer frame { (Fig.~\ref{fig:Trudeau2019}, center and right panels)}. The combination of the {\it Spitzer} contours and our CO observations shows that the MIPS infrared emission does not come from the BCG itself, but mainly from the southern CO source, as previously suggested  by \citet{Webb2015b}. 

The southern galaxy and its tail, which were both detected by us in CO and in continuum, are also spatially aligned at the start of a long chain of of clumpy sources, which originates from the BCG system and broadens toward the southwest.
\citet{Webb2015b} suggest that the stream, which is clearly visible by {\it HST} in the NIR (rest frame UV and optical), is associated with a tidal tail  that originates within the stellar envelope of the BCG from a less massive spectroscopically confirmed merging system.
Our observations now confirm that the tip of the stream is linked to a gas rich companion of the BCG, which is possibly infalling along the stream into the cluster core and thus the BCG itself. {In this case, the infall should occur with a relative velocity that is almost perfectly in the plane of the sky since the line-of-sight velocity is negligible (Figs.~\ref{fig:maps}b, \ref{fig:CO43spectrum}) when compared to the cluster velocity dispersion, which is on the order of several 100s~km/s for massive clusters.}   The cospatial dust continuum and CO tails strengthen our interpretation and suggest that the high intra-cluster medium (ICM) density $\rho_{\rm ICM}$ and high source velocity $v$, as often observed in cluster cores, may favor ram pressure $p\sim \rho v^2$ to strip off the dense CO gas and surrounding dust out of the southern BCG companion, meanwhile boosting its star formation. 
{It is worth mentioning that we cannot firmly exclude the possibility that the tail is instead due to tidal interactions. However, since we only detected one extended tail, we tend to favor the ram-pressure scenario.}
Therefore, the southern BCG satellite is possibly the first high-$z$ analog of 
local jellyfish galaxies \citep[e.g.,][]{Jachym2014,Jachym2017,Jachym2019,George2018,Vulcani2018,Poggianti2019} as well as of 
intermediate-$z$ ones, such as those recently discovered in the clusters Abell~1758 \citep[$z=0.27$,][]{Kalita_Ebeling2019} and CGr32 \citep[$z=0.73$,][]{Boselli2019}.


We used the H$_2$ masses of the northern and southern BCG companions, reported in Sect.~\ref{sec:result_MH2}, to estimate their stellar mass. Under the assumption that the H$_2$-to-stellar mass ratio of the two sources reflects that of field galaxies, we can express the ratio as $M_{{\rm H}_2}/M_\star\sim0.1\,(1+z)^2$ \citep{Carilli_Walter2013}. 
This yields 
${M_\star\sim9.5\times10^{10}M_\odot}$ and ${M_\star\sim1.4\times10^{11}M_\odot}$ for the northern and southern BCG satellites, respectively, that we detected separately in CO.
{ Both molecular gas and} stellar components of these BCG companions are, therefore, non-negligible and 
${\sim0.2}$ and ${\sim0.3}$ of the total $M_\star=4.5\times10^{11}~M_\odot$ of the BCG \citep{Webb2015a}, respectively.

{ By combining the $M_{\rm H_2}$ upper limit for the BCG with its stellar mass estimate, we inferred a low molecular gas content $M_{\rm H_2}/M_\star<0.02$. This is among the lowest gas-mass-to-stellar-mass ratios estimated for distant $z>1$ ellipticals \citep[e.g.,][]{Sargent2015,Gobat2018,Bezanson2019}.}

Our observations also imply an H$_2$-to-dust mass ratio of $M_{{\rm H_2}}/M_d{\simeq455}$ and ${\gtrsim973}$ for the southern and northern BCG companions, which are higher than the values of $\sim100$, typically found for distant star-forming galaxies (see Appendix~\ref{app:formulas}).{ Assuming a lower CO-to-H$_2$ conversion factor than the Galactic one used in this work would alleviate the tension. However, for the northern companion, even in assuming an $\alpha_{\rm CO}=0.8~M_\odot~({\rm K}~{\rm km}~{\rm s}^{-1}~{\rm pc}^2)^{-1}$, which is more typical of starbursts, we would still have a high $M_{{\rm H_2}}/M_d\gtrsim179$.} In order to explain the large $M_{{\rm H_2}}/M_d$ ratio, it could be that a substantial fraction of the observed gas reservoir has been freshly accreted gas by these satellites.

The proximity of massive star-forming satellites, both in velocity, within $\pm150$~km~s$^{-1}$, and in projection, within $\sim20$~kpc from the BCG, suggests that these sources are close to the interaction with the BCG itself. This scenario  implies that the companions will merge into the BCG, which is consistent with previous studies of some distant BCGs \citep{Tremblay2014,Bonaventura2017}. The companions  will thus either provide a significant gas supply, triggering the star formation in the BCG via gas rich (major) mergers,  or they will rejuvenate the BCG stellar population with newly born stars.

The present study is thus one of the first works to observe rapid star formation, $\tau_{\rm dep}=M_{{\rm H}_2}/{\rm SFR}\simeq0.4$~Gyr, at play in a very distant cluster core.
{ The depletion timescale $\tau_{\rm dep}$ was derived assuming the total $M_{{\rm H}_2}=3.3\times10^{11}~M_\odot$ reported in Sect.~\ref{sec:result_MH2} and the  infrared-based ${\rm SFR}=860~M_\odot$/yr inferred by \citet{Webb2015b} for the BCG system, which remains unresolved by {\it Spitzer} MIPS (Fig.~\ref{fig:Trudeau2019}).}
The presence of gas rich and massive star-forming companions of the BCG also shows that we are currently observing the build-up  of the core of the distant SpARCS1049+56 cluster
at the early $z\sim2$ epoch of cluster formation \citep[e.g.,][for a similar case]{Wang2016}.
Thus, it is likely that we are probing the cluster core in a transitioning phase of the cluster formation and possibly the reversal of the  star formation versus density relation \citep[][]{Dressler1980,Kauffmann2004}.
According to this scenario, we speculate that SpARCS1049+56 belongs to the transitioning population of clusters with starbursting cores, which has yet to be explored, thus constituting the link between $z\gtrsim2$ proto-clusters that are still assembling \citep[e.g.,][]{Hayashino2004,Umehata2019}, extending over several megaparsecs, and virialized lower-$z$ clusters, whose cores are typically populated by read-and-dead ellipticals and lenticular galaxies \citep{Oemler1974,Davis_Geller1976}.
The synergy of ongoing and forthcoming observational radio (ALMA, SKA) and optical-infrared facilities (DES, LSST, {\it Euclid}, {\it JWST}) will provide large samples of well characterized starbursting cluster cores and BCGs at the epoch of cluster formation, such as SpARCS1049+56, thus revolutionizing the field of galaxy evolution in high-$z$ dense environments.

\begin{acknowledgements}
We thank the anonymous referee for helpful comments which contributed to improve the paper.
This work is based on observations carried out under project number S18CL with the IRAM NOEMA Interferometer. IRAM is supported by INSU/CNRS (France), MPG (Germany) and IGN (Spain).
GC acknowledges financial support from both the Swiss National Science Foundation (SNSF) and the Swiss Society for Astrophysics and Astronomy (SSAA). GC acknowledges the hospitality of the Observatory of Paris between Oct 28 and Nov 8, 2019 during which the present work has been finalized.
GC thanks Emanuele Daddi for inspiring discussion.
\end{acknowledgements}


\label{lastpage}
\onecolumn
\begin{appendix}
\section{Molecular gas and dust masses}\label{app:formulas}
In this appendix, we provide the formulas adopted in the text to estimate H$_2$ molecular gas and dust masses, or their upper limits.

\subsection{Molecular gas mass from CO(4$\rightarrow$3)}
In the presence of a CO(J$\rightarrow$J-1) emission line, the velocity integrated flux $S_{\rm CO(J\rightarrow J-1)}\Delta v$ is first converted into a velocity integrated luminosity $L^\prime_{\rm CO(J\rightarrow J-1)}$, using Eq.~(3) of \citet{Solomon_VandenBout2005}:
\begin{equation}
\label{eq:LpCO}
L^{\prime}_{\rm CO(J\rightarrow J-1)}=3.25\times10^7\,\frac{S_{\rm CO(J \rightarrow J-1)}\,\Delta\varv}{{\rm Jy}~{\rm km}~{\rm s}^{-1}}\,\bigg(\frac{\nu_{\rm obs}}{\rm GHz}\bigg)^{-2}\,\bigg(\frac{D_L}{\rm Mpc}\bigg)^2\,(1+z)^{-3}\;{\rm K}~{\rm km}~{\rm s}^{-1}~{\rm pc}^2,
\end{equation}
where $\nu_{\rm obs}$  is the observer-frame frequency of the CO(J$\rightarrow$J-1) transition, J is a positive integer denoting the total angular momentum, and $D_L$ is the luminosity distance at the redshift $z$ of the source. The total molecular gas mass can then be estimated as: 
\begin{eqnarray}
\label{eq:MH2}
 M({\rm H_2}) & =\alpha_{\rm CO}L^{\prime}_{\rm CO(1\rightarrow0)}=\frac{\alpha_{\rm CO}L^{\prime}_{{\rm CO}(J\rightarrow J-1)}}{r_{J1}}\,,
\end{eqnarray}
where $r_{J1}= L^{\prime}_{{\rm CO}(J\rightarrow
  J-1)}/L^{\prime}_{{\rm CO}(1\rightarrow0)}$ is the excitation ratio. Typical values for J~$=4$ and J~$=2$ are $r_{41}\simeq0.46$ and $r_{21}\simeq0.85$, respectively, for submillimeter galaxies \citep[][]{Papadopoulos2000,Carilli_Walter2013}.

\subsection{Continuum-based dust mass}
We denote $S_{\nu_{\rm obs}}$ as the observed continuum flux density of a given source at an observed frequency $\nu_{\rm obs}$. The frequency $\nu_{\rm obs}$ is related to the rest-frame frequency $\nu_{\rm rf}$ by the standard relation $\nu_{\rm obs}=\nu_{\rm rf}/(1+z)$.
Similar to previous studies \citep[e.g.,][]{Beelen2006}, $S_{\nu_{\rm obs}}$ can be expressed as a function of the dust mass $M_{d}$ and temperature $T_{d}$, as follows:
\begin{equation}
\label{eq:flux_Mdust}
 S_{\nu_{\rm obs}} = \frac{1+z}{D_L^2}\,M_{d}\,\kappa_d(\nu_{\rm rf})\,B_{\nu_{\rm rf}}(\nu_{\rm rf},T_d)\;,
\end{equation}
where $\kappa_d(\nu)$ is the dust opacity per unit mass of dust and $B_{\nu}(\nu,T)$ is the spectral radiance of a black body of temperature $T$ at frequency $\nu$. It holds:\begin{equation}
 B_\nu(\nu,T) = \frac{2h\,\nu^3}{c^2}\frac{1}{e^{\frac{h\nu}{k_BT}}-1}\;,
\end{equation}
where $h$ is the Planck constant, $c$ is the speed of light, and $k_B$ is the Boltzmann constant.  
The dust opacity is modeled as:
\begin{equation}
 \kappa_d(\nu)=\kappa_0\bigg(\frac{\nu}{250~{\rm GHz}}\bigg)^\beta, 
\end{equation}
with $\kappa_0=0.4$~cm$^{2}$~g$^{-1}$ \citep[][and references therein]{Beelen2006,Alton2004}. We also chose the value $\beta=1.8$ \citep[][]{Scoville2017}, which corresponds to the one that was determined for our Galaxy from {\it Planck}, \citep{PlanckCollaboration2011a,PlanckCollaboration2011b}.
Then, Eq.~\ref{eq:flux_Mdust} can be expressed in a more compact form as:
\begin{eqnarray}
\label{eq:Mdust}
\begin{aligned}
 S_{\nu_{\rm obs}} & =  1.12\frac{M_d}{10^{8}M_\odot}(1+z)^{3+\beta}\,\frac{\kappa_0}{0.4~{\rm cm}^2~{\rm g}^{-1}}
 \bigg(\frac{{\rm 353~GHz}}{250~{\rm GHz}}\bigg)^\beta \bigg(\frac{\nu_{\rm obs}}{353~{\rm GHz}}\bigg)^{2+\beta} \,\bigg(\frac{D_L}{\rm Gpc}\bigg)^{-2}\frac{x}{e^x-1}\cdot\frac{T_d}{35~{\rm K}}\,{
\rm mJy},
 \\
{\rm with}\;  x & = 0.484\frac{35~{\rm K}}{T_d}\frac{\nu_{\rm obs}}{353~{\rm GHz}}(1+z).
\end{aligned}
\end{eqnarray}
By inverting the last equation, the dust mass can be derived once the continuum flux density $S_{\nu_{\rm obs}}$ is known.

\subsection{Continuum-based molecular mass}
Under some assumptions, continuum-based molecular mass can be estimated, similarly to $M_d$.
Following \citet{Scoville2014}, we assume a direct proportionality between the continuum flux density $S_{\nu_{\rm obs}}$ in the Rayleigh-Jeans regime and the interstellar medium (ISM) mass density, that is:
\begin{equation}
\label{eq:kappaISM}
\kappa_{\rm ISM}(\nu) = \kappa_d(\nu)\frac{\rho_d}{\rho_{\rm ISM}}\,,
\end{equation}
where $\rho_d$ and $\rho_{\rm ISM}$ are the dust and interstellar medium (ISM) mass densities, respectively, while $\kappa_{\rm ISM}(\nu)$ is the dust  opacity per unit ISM mass. 
As found by \citet{Scoville2014} for distant star-forming galaxies out to $z=2,$ the continuum flux density $S_{\nu_{\rm obs}}$  can then be expressed as:
\begin{eqnarray}
\begin{aligned}
\label{eq:Mmol_continuum}
S_{\nu_{\rm obs}}& =1.13\frac{M_{{\rm H}_2}}{10^{10}M_\odot}(1+z)^{3+\beta}\,\bigg(\frac{\nu_{\rm obs}}{353~{\rm GHz}}\bigg)^{2+\beta}\,\bigg(\frac{D_L}{\rm Gpc}\bigg)^{-2}\frac{x}{e^x-1}\,{\rm mJy}.\\
\end{aligned}
\end{eqnarray}
The last estimate assumes that the total ISM mass is equal to $1.36\,M_{{\rm H}_2}$ in order to account for the contribution of heavy elements (mostly He) to the ISM \citep{Scoville2014,Scoville2016}.
By inverting Eq.~\ref{eq:Mmol_continuum}, the molecular gas mass can be estimated once the continuum flux density $S_{\nu_{\rm obs}}$ is known. A comparison between Eq.~\ref{eq:Mdust} and Eq.~\ref{eq:Mmol_continuum} yields typical H$_2$-to-dust mass ratios on the order of $\sim100$.

\end{appendix}


\begin{thebibliography}{99}
\bibitem[Alton et al.(2004)]{Alton2004} Alton, P.~B., Xilouris, E.~M., Misiriotis, A. et al., 2004, A\&A, 425, 109
\bibitem[Andreon \& Congdon(2014)]{Andreon_Congdon2014} Andreon, S. \& Congdon, P. 2014, A\&A, 568, 2
\bibitem[Beelen et al.(2006)]{Beelen2006} Beelen, A., Cox, P., Benford, D.~J. et al. 2006, ApJ, 642, 694 
\bibitem[Berta et al.(2016)]{Berta2016} Berta S., Lutz D., Genzel R. et al. 2016, A\&A, 587, A73
\bibitem[Bezanson et al.(2019)]{Bezanson2019} { Bezanson, R., Spilker, J.,  Williams, C.~C. et al. 2019, ApJ, 873, 19}
\bibitem[Bonaventura et al.(2017)]{Bonaventura2017} Bonaventura, N.~R., Webb, T.~M.~A., Muzzin, A. et al. 2017, MNRAS, 469, 1259
\bibitem[Boselli et al.(2019)]{Boselli2019} Boselli A., Epinat B., Contini T. et al. 2019, arXiv190905491
\bibitem[Brodwin et al.(2012)]{Brodwin2012} Brodwin, M., Gonzalez, A.~H., Stanford, S.~A. et al. 2012, ApJ, 753, 162
\bibitem[Carilli \& Walter(2013)]{Carilli_Walter2013} Carilli, C.~L., \& Walter, F. 2013, ARA\&A, 51, 105
\bibitem[Castignani et al.(2019)]{Castignani2019} Castignani, G., Combes, F., Salom\'e, P. et al. 2019, A\&A, 623, 48
\bibitem[Collins et al.(2009)]{Collins2009} Collins, C.~A., Stott, J.~P.,  Hilton, M. et al. 2009, Nature, 458, 603
\bibitem[Coogan et al.(2018)]{Coogan2018} Coogan, R.~T., Daddi, E., Sargent, M.~T. et al. 2018, MNRAS, 479, 703
\bibitem[Davis \& Geller(1976)]{Davis_Geller1976} Davis, M. \& Geller, M.~J. 1976, ApJ, 208, 13
\bibitem[De Lucia \& Blaizot(2007)]{DeLucia_Blaizot2007} De Lucia, G., \& Blaizot, J. 2007, MNRAS, 375, 2
\bibitem[Dressler(1980)]{Dressler1980} Dressler A., 1980, ApJ, 236, 351
\bibitem[Edge(2001)]{Edge2001} Edge, A. C., 2001, MNRAS, 328, 762
\bibitem[Emonts et al.(2013)]{Emonts2013} Emonts, B.~H.~C., Feain, I., R\"{o}ttgering, H.~J.~A. et al. 2013, MNRAS, 430, 3465
\bibitem[Emonts et al.(2016)]{Emonts2016} Emonts, B.~H.~C., Lehnert, M.~D., Villar-Martín, M. et al. 2016, Science, 354, 1128
\bibitem[Fabian et al.(2006)]{Fabian2006} Fabian, A.~C., Sanders, J.~S., Taylor, G.~B. et al., 2006, MNRAS, 366, 417
\bibitem[Finner et al.(2020)]{Finner2020} Finner, K., Jee, M.~J., Webb, T., et al. 2020, arXiv:2002.01956
\bibitem[Fogarty et al.(2019)]{Fogarty2019} Fogarty, K., Postman, M., Li, Y. et al. 2019, ApJ, 879, 103
\bibitem[Fraser-McKelvie et al.(2014)]{FraserMcKelvie2014} Fraser-McKelvie, A., Brown, M.~J.~I., \& Pimbblet, K.~A., 2014, MNRAS, 444, 63
\bibitem[George et al.(2018)]{George2018} George, K., Poggianti, B.~M., Gullieuszik, M. et al. 2018, MNRAS, 479, 4126
\bibitem[Gobat et al.(2013)]{Gobat2013} Gobat, R.; Strazzullo, V.; Daddi, E. et al. 2013, ApJ, 776, 9
\bibitem[Gobat et al.(2018)]{Gobat2018} { Gobat, R., Daddi, E., Magdis, G., et al. 2018, Nature Astronomy, 2, 239}
\bibitem[Hamer et al.(2012)]{Hamer2012} Hamer, S.~L.,  Edge, A.~C., Swinbank, A.~M. et al. 2012, MNRAS, 421, 3409 
\bibitem[Hayashino et al.(2004)]{Hayashino2004} Hayashino, T., Matsuda, Y., Tamura, H. et al. 2004, AJ, 128, 2073
\bibitem[J\'{a}chym et al.(2014)]{Jachym2014} J\'{a}chym, P., Combes, F., Cortese, L. et al. 2014, ApJ, 792, 11 
\bibitem[J\'{a}chym et al.(2017)]{Jachym2017} J\'{a}chym, P., Sun, M., Kenney, J.~D.~P. et al. 2017, ApJ, 839, 114 
\bibitem[J\'{a}chym et al.(2019)]{Jachym2019} J\'{a}chym, P., Kenney, J.~D.~P., Sun, M. et al. 2019, ApJ, 883, 145
\bibitem[Kalita \& Ebeling(2019)]{Kalita_Ebeling2019} Kalita, B.~S. \& Ebeling, H. 2019, arXiv191011898
\bibitem[Kauffmann et al.(2004)]{Kauffmann2004} Kauffmann, G., White, S.~D.~M., Heckman, T.~M. et al. 2004, MNRAS, 353, 713
\bibitem[Lauer et al.(2014)]{Lauer2014} Lauer, T.~R., Postman, M., Strauss, M.~A. et al. 2014, ApJ, 797, 82
\bibitem[Lidman et al.(2012)]{Lidman2012} Lidman, C., Suherli, J., Muzzin, A., et al. 2012, MNRAS, 427, 550
\bibitem[McDonald et al.(2014)]{McDonald2014} McDonald, M., Swinbank, M., Edge, A.~C. et al. 2014, ApJ, 784, 18
\bibitem[McDonald et al.(2016)]{McDonald2016} McDonald, M., Stalder, B., Bayliss, M., et al. 2016, ApJ, 817, 86
\bibitem[McNamara et al.(2014)]{McNamara2014} McNamara, B.~R., Russell, H.~R., Nulsen, P.~E.~J. et al. 2014, ApJ, 785, 44 
\bibitem[Newman et al.(2014)]{Newman2014} Newman, A.~B., Ellis, R.~S., Andreon, S., et al. 2014, ApJ,788, 51
\bibitem[Oemler et al.(1974)]{Oemler1974} Oemler, A.~Jr. 1974, ApJ, 194, 1O 
\bibitem[Olivares et al.(2019)]{Olivares2019} Olivares, V., Salom\'{e}, P., Combes, F., et al. 2019, A\&A, 631, 22O
\bibitem[Papadopoulos et al.(2000)]{Papadopoulos2000} Papadopoulos, P.~P., R\"{o}ttgering, H.~J.~A., van der Werf, P.~P. et al. 2000, ApJ, 528, 626
\bibitem[Planck Collaboration(2011a)]{PlanckCollaboration2011a} Planck Collaboration early results XXI, 2011a, A\&A, 536, A21
\bibitem[Planck Collaboration(2011b)]{PlanckCollaboration2011b} Planck Collaboration early results XXV, 2011b, A\&A, 536, A25
\bibitem[Planck Collaboration(2018)]{PlanckCollaborationVI2018} Planck Collaboration results VI, 2018, arXiv:1807.06209
\bibitem[Poggianti et al.(2019)]{Poggianti2019}  Poggianti, B.~M., Ignesti, A., Gitti, M. et al. 2019, arXiv191011622
\bibitem[Reynolds \& Fabian(1996)]{Reynolds_Fabian1996} Reynolds, C.~S., \& Fabian, A.~C. 1996, MNRAS, 278, 479
\bibitem[Riess et al.(2019)]{Riess2019} Riess, A.~G., Casertano, S., Yuan, W. et al. 2019, ApJ, 876, 85
\bibitem[Russell et al.(2014)]{Russell2014} Russell, H.~R., McNamara, B.~R., Edge, A.~C. et al. 2014,  ApJ, 784, 78 
\bibitem[Russell et al.(2019)]{Russell2019} Russell, H~R.; McNamara, B.~R., Fabian, A.~C. et al. 2019, MNRAS, 490, 3025
\bibitem[Salom\'{e} \& Combes(2003)]{Salome_Combes2003} Salom\'{e} P. \&  Combes, F. 2003, A\&A, 412, 657
\bibitem[Salom\'{e} et al.(2006)]{Salome2006} Salom\'{e} P., Combes, F., Edge, A.~C. et al. 2006, A\&A, 454, 437
\bibitem[Sargent et al.(2015)]{Sargent2015} Sargent, M. T., Daddi, E., Bournaud, F., et al. 2015, ApJL, 806, L20
\bibitem[Scoville et al.(2014)]{Scoville2014} Scoville, N., Aussel, Sheth, K. et al. 2014, ApJ, 783, 
\bibitem[Scoville et al.(2016)]{Scoville2016} Scoville, N.,  Sheth, K.,  Aussel, H. et al. 2016, ApJ,820, 83, 2016
\bibitem[Scoville et al.(2017)]{Scoville2017} Scoville, N., Lee, N., Vanden Bout, P., et al. 2017, 837, 150 
\bibitem[Solomon \& Vanden Bout(2005)]{Solomon_VandenBout2005} Solomon, P.~M. \& Vanden Bout, P.~A., 2005, ARA\&A, 43, 677
\bibitem[Stanford et al.(2012)]{Stanford2012} Stanford, S.~A., Brodwin, M., Gonzalez, A.~H., et al. 2012, ApJ, 753, 164
\bibitem[Stott et al.(2011)]{Stott2011} Stott, J.~P., Collins, C.~A., Burke, C. et al., 2011, MNRAS, 414, 445
\bibitem[Strazzullo et al.(2018)]{Strazzullo2018} Strazzullo, V.; Coogan, R. T., Daddi, E., et al. 2018, ApJ, 862, 64
\bibitem[Tremblay et al.(2014)]{Tremblay2014} Tremblay, G.~R., Gladders, M.~D., Baum, S.~A. et al. 2014, ApJ, 790, 26
\bibitem[Tremblay et al.(2016)]{Tremblay2016} Tremblay, G.~R., Oonk, J.~B.~R., Combes, F. et al. 2016, Nature, 534, 218
\bibitem[Trudeau et al.(2019)]{Trudeau2019} Trudeau, A., Webb, T., Hlavacek-Larrondo, J. et al. 2019, MNRAS, 487, 1210
\bibitem[Umehata et al.(2019)]{Umehata2019} Umehata H., Fumagalli M., Smail I., et al. 2019, Science, 366, 97
\bibitem[Vulcani et al.(2018)]{Vulcani2018} Vulcani, B., Poggianti, B.~M.. Gullieuszik, M. et al. 2018, ApJ, 866, 25
\bibitem[Wang et al.(2016)]{Wang2016} Wang T., Elbaz D., Daddi E. et al. 2016, ApJ,828, 56
\bibitem[Webb et al.(2015a)]{Webb2015a} Webb, T., Muzzin, A., Noble, A. et al. 2015a, ApJ, 814, 96 
\bibitem[Webb et al.(2015b)]{Webb2015b} Webb, T., Noble, A., De Groot, A. et al. 2015b,  ApJ, 809, 173
\bibitem[Webb et al.(2017)]{Webb2017} Webb, T., Lowenthal, J., Yun, M. et al. 2017, ApJL, 844, 17
\bibitem[Zhang et al.(2016)]{Zhang2016} Zhang, Y., Miller, C., McKay, T., et al. 2016, ApJ, 816, 98
\bibitem[Zirbel(1996)]{Zirbel1996} Zirbel, E.~L. 1996, ApJ, 473, 713




\end{thebibliography}
\end{document}